\documentclass[a4paper]{article}
\usepackage{bm}
\usepackage{amsmath,amssymb}
\usepackage{graphicx}
\usepackage{url}
\newcommand{\J}{{\rm J}}
\renewcommand{\H}{{\rm H}}
\newcommand{\cs}{{\rm cs}}

\renewcommand{\Im}{{\rm Im\,}}
\title{Guiding of Light with Pinholes}
\author{Makoto Morinaga\\
Institute for Laser Science, University of Electro-Communications\\
Chofu, Tokyo, 182-8585 JAPAN}
\begin{document}
\maketitle

\begin{abstract}
A new type of light waveguide using linearly aligned pinholes
is presented. Results of basic experiments are compared with
theoretical estimates calculated using continuous model.
Since this waveguide utilizes no transparent material,
it can be used to guide electromagnetic waves of wide wavelength
ranges as well as other waves such as matter waves.
\end{abstract}


\section{Introduction}
\begin{figure}[htbp]
\includegraphics[width=8.5cm]{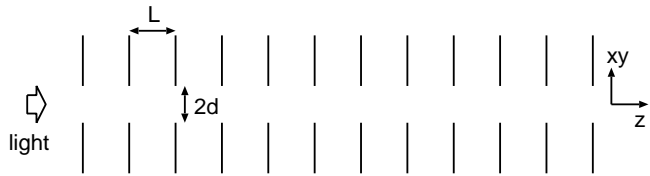}
\caption{Construction of the pinhole waveguide. Pinholes of
a diameter $d$ are aligned on a straight line with spacing $L$
between the pinholes.}
\label{scheme}
\end{figure}

Optical fibers are extremely useful as light waveguides
and are widely used not only for
classical communications\cite{FOC}
but also for quantum communications\cite{QCC}.
Their transmission loss can be especially low for some specific
wavelength such as $\lambda=1.55\mu$m where the absorption of
silica glass is at minimum. Conversely, one needs a good
transparent material to obtain a good transmissivity.
In this paper, we propose a new type of waveguide composed
of linearly aligned pinholes of same diameter (fig.\ref{scheme}).
Since no transparent material is required for its construction,
it can be used
to guide electromagnetic fields of wide frequency ranges, or
waves of other kinds such as matter waves. Or it could also
be used to guide and/or confine atoms with light guided
by this pinhole waveguide because the space where light
passes through is vacant.
Such a structure is meaningless in geometrical optics since
whether a ray transmits through this structure depends only
on the geometrical arrangement of the ray and the first and the
last pinholes, and pinholes in between play no role for the
transmission.
However, if we treat light as wave, as described in the appendix
(see also \cite{tbp}), transmission loss at
each pinhole has a form $\beta\left(\frac {\lambda L}{d^2}\right)
^{\frac 32}$
where $\lambda$ is the wavelength of the light, $L$ is the spacing
between pinholes, $d$ is the radius of the pinholes, and $\beta$
is a constant that depends on the order of the transverse mode
of the light (see fig.\ref{scheme}).
Thus the transmission loss per unit length is proportional to
$\sqrt L$ so that it can in principle be made arbitrary small
by taking the spacing $L$ between the pinholes smaller and smaller.
In the following, we present basic experiments
with pinhole waveguides and, in the appendix, we will outline a
theoretical treatment of this waveguide using continuous
model to analyze the experimental results.

\section{Experiments with pinhole waveguides}
\begin{figure}[htbp]
 \includegraphics[width=8.0cm]{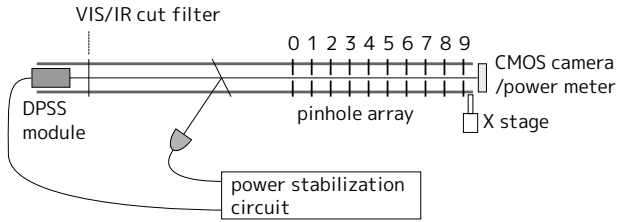}\\
\caption{Light from a DPSS laser module propagates through pinholes and
is detected by either a CMOS image sensor or a power meter
at the end of the pinhole array which also can be inserted
in between the pinholes.}
\label{setup_d_}
\end{figure}

The experimental setup is schematically shown in fig.\ref{setup_d_}.
A DPSS (Diode Pumped Solid State)
laser module generates TEM${}_{00}$ output of wavelength
at both 1064nm and 532nm. 1064nm (532nm) wavelength is selected by
inserting a VIS cut filter (IR cut filter). Part of the laser beam
is reflected into a photo diode for power stabilization.
When the laser beam enters the pinhole array, the beam size is
considerably larger than the size of the pinhole and we can
regard the incident wave as a plane wave.
Each pinhole is mounted on a $xy$-translation stage which is fixed
on a linear rail lying in $z$-direction. Up to 10 pinholes can be set
on the rail with minimum spacing of 15mm (29mm to insert the
image sensor or the power meter between the pinholes).
\subsection{Pinhole alignment}
\begin{figure}[htbp]
 \includegraphics[width=7cm]{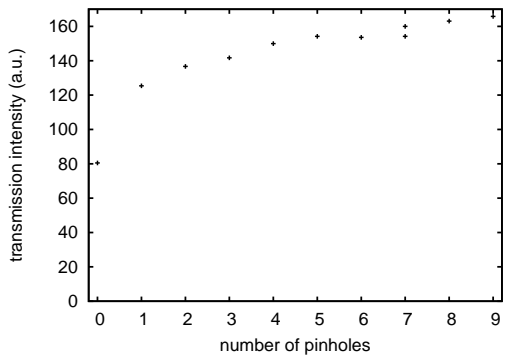}
\caption{The power of the output of the pinhole array is
plotted while
inserting pinholes one by one (see text).
Pinhole diameter: $2d=$0.5mm. Light wavelength: $\lambda=$532nm.}
\label{tmt}
\end{figure}

The alignment procedure of the pinholes on a straight line is as follows.
The power meter is always set at the end of the pinhole array during
this procedure and the light at 532nm wavelength is used.
First we set no pinhole except the last pinhole
(pinhole no.9 in fig.\ref{setup_d_})
and maximize the power by adjusting the $xy$-position of this pinhole
(the power is plotted as '0' in the horizontal axis of fig.\ref{tmt}).
Then we set the first pinhole (pinhole no.0 in fig.\ref{setup_d_})
and maximize the power by adjusting the $xy$-position of this pinhole
(plotted as '1' in the horizontal axis of fig.\ref{tmt}).
And then pinhole no.1 (plotted as '2'), pinhole no.2 (plotted as '3'),
and so on. The spacing between the pinholes is $L=45$mm.
From fig.\ref{tmt} we see that the transmitted light power increases
with increasing number of pinholes, which cannot be explained by
geometrical optics.
\subsection{Propagation through the pinhole array}
\begin{figure}[htbp]
 \includegraphics[width=4.8cm]{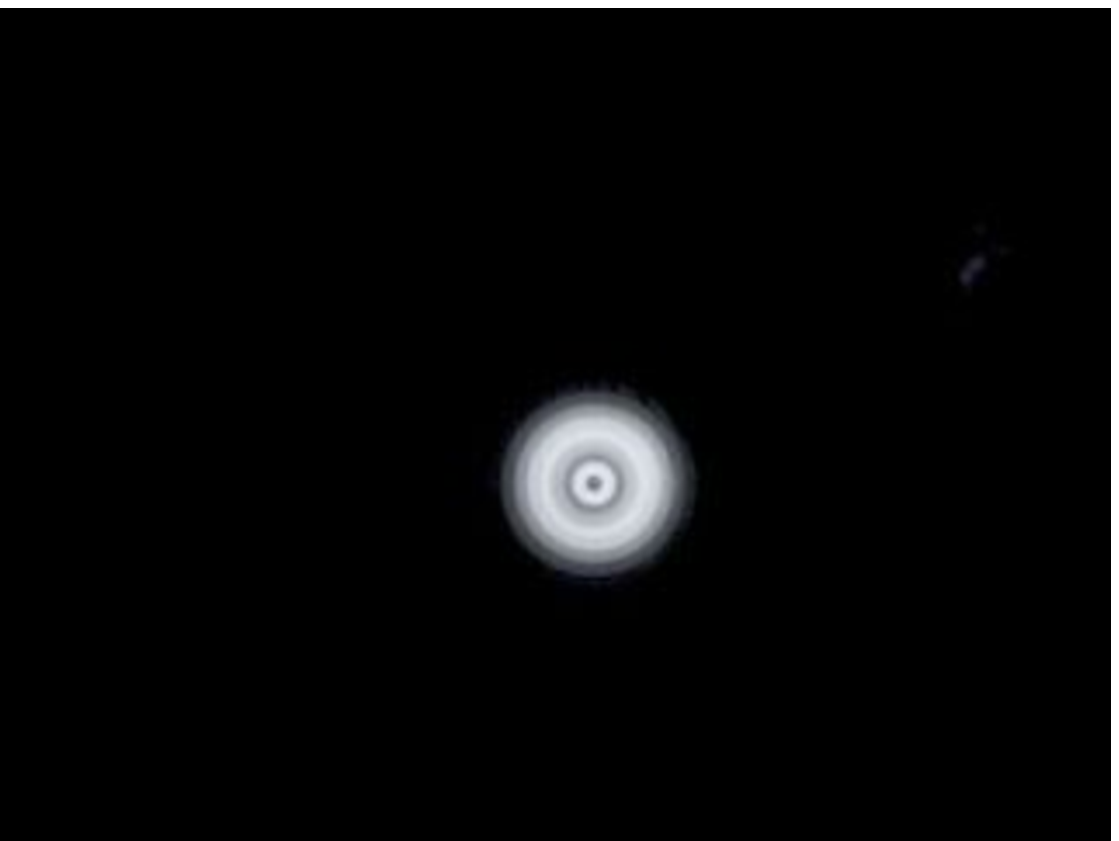}\hspace{2mm}
 \includegraphics[width=4.8cm]{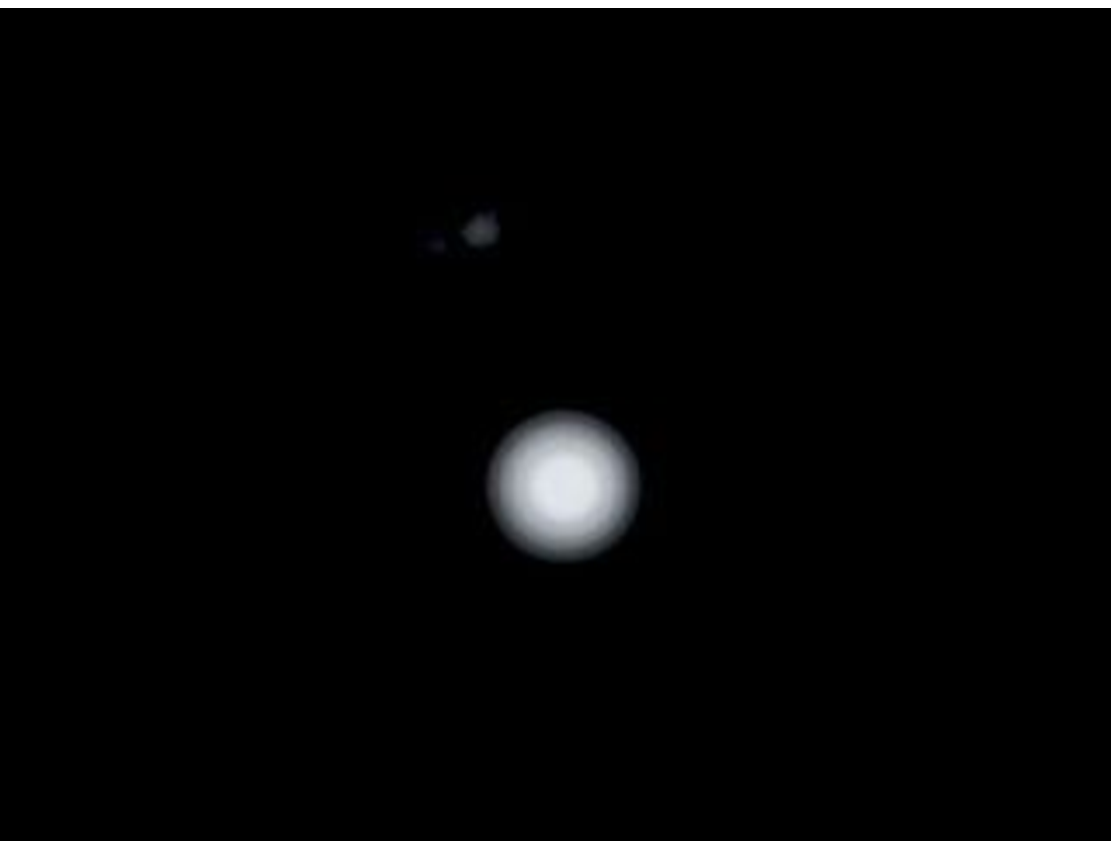}\\
\caption{Beam image after pinhole no.0 (left) and pinhole no.3 (right).}
\label{after}
\end{figure}

After aligning all the 10 pinholes light power after each pinhole
is measured and compared with the theoretical curve calculated
using the equation (\ref{Pz}) in the appendix (fig.\ref{decay}).
The theoretical curves are plotted with no fitting parameter
except that initial power is normalized to 1 for both experiment and theory.
The tendency that the experimental value is lower than the
theoretical curve might be explained by the fact that the misalignment
of the pinhole always decreases the power from that without misalignment.
The initial square beam cut out from the incident plane wave
contains high order transverse modes that attenuate fast compared
with the lowest order mode leading to the initial rapid decay.
After propagating through several pinholes, lowest order mode
dominates then showing slower decay.
\begin{figure}[htbp]
 \includegraphics[width=7cm]{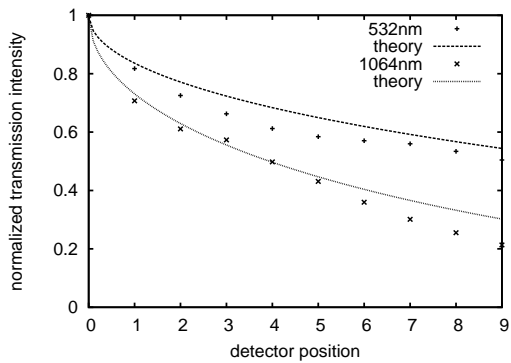}
\caption{Light power after each pinhole is plotted.
The power after the pinhole no.0 is normalized to 1.
Pinhole spacing: $L$=45mm. $2d=$0.5mm.
Two lines are theoretical curves calculated using the
continuous model.}
\label{decay}
\end{figure}
In fig.\ref{after} beam images after the pinhole no.0 and no.3 are
shown ($\lambda=$532nm). The latter shows smooth profile with a peak
intensity at the
center which qualitatively confirms the explanation given above
(the distance of about 17mm from the pinhole to the image sensor
makes fine structure in the left image due to diffraction).
\subsection{Bending of light}
\begin{figure}[htbp]
 \includegraphics[width=7.5cm]{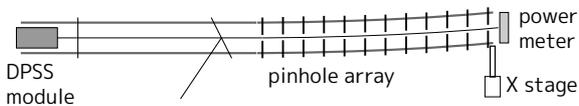}\\
\caption{Schematics of the light bending experiment (see text).}
\label{bend_d_}
\end{figure}

\begin{figure}[htbp]
 \includegraphics[width=7.5cm]{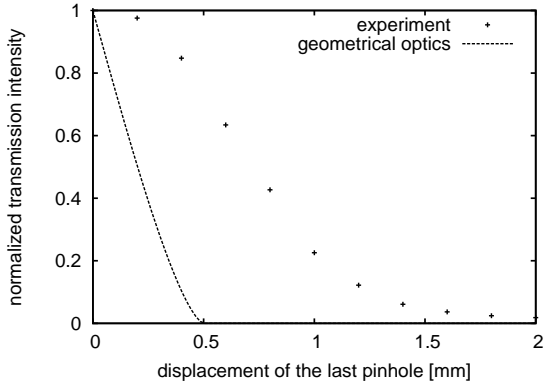}\\
\caption{Power of the transmitted light is measured while bending
the waveguide.
$2d=$0.5mm. $\lambda=$532nm. $L=$45mm.
Dotted line is the theoretical curve following the
geometrical optics.}
\label{bendnn}
\end{figure}

The linear rail on which pinholes is sitting is fixed to the
optical table at three points: at the
left end, in the middle near the pinhole no.0, and at the
right end. We remove the fixing screw at the right end and
push this end to the transverse direction. Then the rail
is bent elastically and the line on which pinholes lies
is curved with a uniform curvature (fig.\ref{bend_d_}).
In fig.\ref{bendnn} we plot the power of the output light
versus the displacement of the last pinhole (pinhole no.9).
The experimental value is compared with a theoretical curve
assuming the geometrical optics.
Certain amount of light is transmitted even with displacements
larger than the diameter of the pinhole (0.5mm).
Note also that the initial decrease seems to be quadratic in
the displacement while
geometrical optics predicts a linear response.
\section{Conclusion and outlook}
In this paper we presented basic experiments of light guiding
with a pinhole array. The results roughly confirm the theoretical
estimates carried out using continuous model. However further
study is needed to understand the details of this new waveguide,
such as how the thickness of the pinholes affects in the
transmission.
\vspace{5mm}

\noindent
{\bf Acknowledgements}\\
This work was partly supported by the
AMADA FOUNDATION and the
Photon Frontier Network Program (MEXT).

\appendix
\section{Continuous model}
\begin{figure}[htbp]
\includegraphics[width=6.5cm]{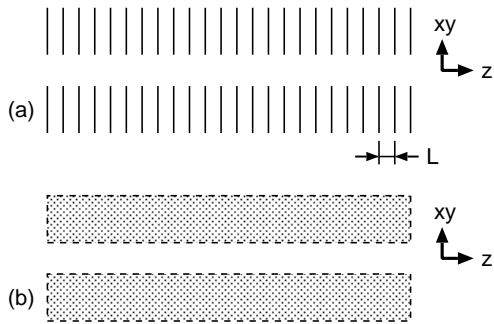}
\caption{(a) Array of masks separated by $L$.
(b) The mask array is replaced with
continuous absorbing medium.}
\label{cam}
\end{figure}

Dealing directly with a discrete array of masks (pinholes, slits,...)
for theoretical analysis is not an easy task.
Instead, we introduce in this appendix a model in which the discrete
set of masks is replaced with continuous medium of some absorbance
that fills the closed region of the masks (see Fig. \ref{cam}).
This continuous model was first introduced for an array of half-planes
to account the enhanced quantum reflection of matter waves from
the ridged surfaces\cite{KO,KO2}.
The lowest order transverse mode function for the slit array
and its loss parameter are
also calculated already\cite{KM}.
Here we will determine all the transverse mode functions and
their propagation parameter for the case of the slit array and
the pinhole array. The light field is treated as a scalar field
(scalar theory).
The wave equation for this model is given by
\[
 -k_0^2\psi(x,y,z)=\{1-i\epsilon\,\theta(x,y)\}\nabla^2\psi(x,y,z)
\]
where $k_0$ is the wavenumber of light,
$\theta(x,y)$ is a step function that takes value 1 (0) in the
closed (open) region 
of masks, and $\epsilon$ is a positive constant related to the
absorptance of the medium. As we shall see below $\epsilon\ll 1$
for our system under consideration.
First we consider a plane wave propagating through the area
filled uniformly with such absorbing medium.
Taking $z$-axis as the direction of propagation,
the wave equation is written as
\[
 -k_0^2\psi(z)=(1-i\epsilon)\psi''(z).
\]
Its solution is given by
\[
 \psi(z)=e^{ik_zz}
\]
with
$k_0^2=(1-i\epsilon)k_z^2
\approx\left\{\left(1+\frac i2\epsilon\right)^{-1}k_z\right\}^2$
so that
$k_z=\left(1+\frac i2\epsilon\right)k_0$ and
$|\psi(z)|^2=e^{-\zeta_0z}$ with the intensity absorptance
$\zeta_0=\epsilon k_0$.
Given that this absorbing medium imitates a stack of opaque
masks separated by a distance $L$, we expect that
$\zeta_0\sim\frac 1L$ so that $\epsilon\sim\frac 1{k_0L}$.
Thus we shall write $\epsilon=\frac\xi{k_0L}$ with a positive parameter $\xi$
of order of 1.
We consider the parameter region where the separation of masks $L$
is much larger than the wavelength, so that $\epsilon\ll 1$.
The continuous model itself cannot determine the value of
$\xi$ (or $\epsilon$). By comparing the attenuation of a wave
propagating in a slit
waveguide calculated using continuous model with that calculated with
direct method\cite{tbp} it is shown that
\begin{equation}
 \xi=\frac 92\pi.
\label{xi}
\end{equation}

\subsection{Slit array}
Consider an array of slits of opening width $2d$
(open for $|x|<d$).
The wave equation is written as
\[
 -k_0^2\psi(x,z)
=\{1-i\epsilon\,\theta(x^2-d^2)\}(\partial_x^2+\partial_z^2)\psi(x,z)
\]
where $\theta$ is the conventional step function defined as
\[
 \theta(s)\equiv\left\{
\begin{array}{ll}
 0&(s<0) \\
 1&(s\ge 0)
\end{array}
\right..
\]
Because of the translational symmetry in
$z$-axis direction, we can assume the form of the solution as
$\psi(x,z)=\varphi(x)\,e^{ik_zz}$
(a general solution is the sum of such solutions).
Below we also assume that the wave propagates nearly in $z$
direction so that $k_z\approx k_0$.

\paragraph{Region A (inside the opening of slits) $|x|\le d$}
The wave equation for the transverse wavefunction $\varphi(x)$ here is
\[
 -(k_0^2-k_z^2)\varphi(x)
=\varphi''(x)
\]
with a solution
\[
 \varphi(x)=\cs(k_Ax),
\]
where $\cs$ is defined as
\begin{equation}
 \cs(u)=\left\{
\begin{array}{cl}
 \cos u&(even\ parity)\\\sin u&(odd\ parity)
\end{array}
\right.,
\label{parity_}
\end{equation} 
and $k_A$ satisfies $k_0^2=k_A^2+k_z^2$.
We are considering the situation in which the wave is nearly confined
in the opening region $|x|\le d$ and hence
$\varphi(\pm d)\approx 0$,
so that $k_Ad\approx \frac{n+1}2\pi\ (n=0,1,2,...)$
provided that we use $\cos$ ($\sin$) in 
(\ref{parity_}) for even (odd) $n$.
Defining $k_n\equiv\frac{n+1}{2d}\pi$ and write $k_A=k_n+\beta+i\gamma$
with real numbers $\beta$ and $\gamma$, then
$|\beta d|\ll 1$ and $|\gamma d|\ll 1$.
\[
 \varphi(\pm d)=\cs(\pm k_A d)
\approx
\pm(\beta+i\gamma)d\,\cs'(\pm k_n d)
\]
\[
 \varphi'(\pm d)=k_A\,\cs'(\pm k_A d)
\approx k_n\,\cs'(\pm k_n d)
\]
\begin{equation}
\left.\frac{\varphi'}\varphi\right|_{x=\pm d}\approx
\pm\frac{k_n}{(\beta+i\gamma)d}
\label{bcA}
\end{equation}
\paragraph{Region B (outside the opening of slits): $|x|\ge d$}
Here the wave equation is
\[
 -\{k_0^2-(1-i\epsilon)k_z^2\}\varphi(x)
=(1-i\epsilon)\varphi''(x)
\]
so that the solution is, taking into account that
it should not diverge when $x\rightarrow\pm\infty$,
\begin{equation}
 \varphi(x)\propto e^{ik_B|x|}
\label{eqB}
\end{equation}
where $k_B$ satisfies $k_0^2=(1-i\epsilon)(k_B^2+k_z^2)$ and
$\Im k_B>0$
(the proportionality factor in (\ref{eqB}) has opposite sign
for $x\ge d$ and $x\le -d$ for the odd parity solution).
Thus
\begin{equation}
\left.\frac{\varphi'}\varphi\right|_{x=\pm d}=\pm ik_B
\label{bcs}
\end{equation}
\paragraph{Boundary condition at $x=\pm d$}
By requiring $\varphi(x)$ and $\varphi'(x)$
is continuous at $x=\pm d$, from (\ref{bcA}) and (\ref{bcs}) we find
\begin{equation}
ik_B
=\frac{k_n}{(\beta+i\gamma)d}.
\label{bcs_}
\end{equation}
By noting $k_B^2-k_A^2=\frac{i\epsilon}{1-i\epsilon}k_0^2$ and
\[
 |k_B|=\frac{|k_n|}{|\beta+i\gamma|d}\gg |k_n|\approx |k_A|,
\]
we see
\begin{equation}
 k_B^2\approx i\epsilon k_0^2
\label{k_B2}
\end{equation}
so that
\begin{equation}
 k_B=\frac{1+i}{\sqrt 2}\sqrt\epsilon k_0.
\label{k_B}
\end{equation}
Using (\ref{bcs_})
\[
 \beta+i\gamma=\frac{k_n}{ik_Bd}=-\frac{1+i}{\sqrt 2}
\frac{k_n}{\sqrt\epsilon k_0d}
\]
and we obtain
\begin{equation}
 k_A=k_n+\beta+i\gamma=\left(1-\frac{1+i}
{\sqrt{2\epsilon}k_0d}\right)k_n.
\label{k_A}
\end{equation}
From the assumption that $|\beta d|\ll 1$ and $|\gamma d|\ll 1$
we see $\sqrt\epsilon k_0d\gg 1$, i.e.
\begin{equation}
 \frac L{k_0d^2}=\frac 1{2\pi}\frac{\lambda L}{d^2}\ll 1.
\label{cp}
\end{equation}
Finally $k_z$ is derived as
\begin{equation}
 k_z=\sqrt{k_0^2-k_A^2}\approx k_0-\frac 12\frac{k_A^2}{k_0}
\approx k_0-\frac{k_n^2}{2k_0}+\frac{k_n^2}
{\sqrt{2\epsilon}k_0^2d}i
\end{equation}
From this we calculate the light attenuation along the waveguide
\[
 |e^{ik_zz}|^2=e^{-\zeta z}
\]
where the attenuation coefficient $\zeta=2\Im k_z$ is
calculated as
\[
 \zeta=\frac{\sqrt 2k_n^2}{\sqrt{\epsilon}k_0^2d}
=(n+1)^2\frac{\sqrt 2\pi^2}{4\sqrt{\epsilon}k_0^2d^3}
=(n+1)^2\frac{\sqrt{2L}\pi^2}{4\sqrt{\xi k_0^3}d^3}
\]
using $\epsilon=\frac\xi{k_0L}$.
The attenuation per length $L$ (i.e. per one slit) $\zeta L$
can be written as a function of a single dimensionless parameter
$\frac{\lambda L}{d^2}$:
\[
 \zeta L
=(n+1)^2\frac{\sqrt{2}\pi^2}{4\sqrt{\xi}}
\left(\frac L{k_0d^2}\right)^{\frac 32}
=(n+1)^2\frac{\sqrt{\pi}}{8\sqrt{\xi}}
\left(\frac{\lambda L}{d^2}\right)^{\frac 32}
\]

\subsection{Pinhole array}
In the case of an array of pinholes of diameter $d$,
using cylindrical coordinate $(r,\phi,z)$,
the wave equation is written as
\[
 -k_0^2\psi(r,\phi,z)
=\{1-i\epsilon\,\theta(r^2-d^2)\}\left(\partial_r^2+\frac 1r\partial_r
+\frac 1{r^2}\partial_{\phi}^2+\partial_z^2\right)\psi(r,\phi,z).
\]
In the same way as in the case of slit array, we shall derive a solution
of the form $\psi(r,\phi,z)=\varphi(r,\phi)e^{ik_zz}$ with
$k_z\approx k_0$.

\paragraph{Region A (inside the opening of pinholes): $|r|\le d$}
The wave equation
\[
 -(k_0^2-k_z^2)\varphi(r,\phi)
=\left(\partial_r^2+\frac 1r\partial_r
+\frac 1{r^2}\partial_{\phi}^2\right)\varphi(r,\phi)
\]
is solved using the Bessel functions of the 1st kind
$J_m\ (m=0,\pm 1,\pm 2,...)$:
\[
 \varphi(r,\phi)=\J_m(k_Ar)e^{im\phi}
\]
where $k_A$ satisfies $k_0^2=k_A^2+k_z^2$.
Again we postulate $\varphi(d,\phi)\approx 0$ which leads to
$k_Ad\approx \rho^{(m)}_n\ (n=0,1,2,...)$.
Here $\rho^{(m)}_0,\,\rho^{(m)}_1,\,\rho^{(m)}_2,...$
are positive zeros of $J_m(\rho)$ sorted in ascending order.
We define
$k^{(m)}_n\equiv\frac{\rho^{(m)}_n}d$
and write $k_A=k^{(m)}_n+\beta+i\gamma$ using real numbers
$\beta$ and $\gamma$ with
$|\beta d|\ll 1$ and $|\gamma d|\ll 1$.
\[
 \varphi(d,\phi)=\J_m(k_A d)e^{im\phi}\approx
(\beta+i\gamma)d\,\J_m'(k_n d)e^{im\phi}
\]
\[
 \partial_r\varphi|_{(d,\phi)}=k_A\,\J_m'(k_Ad)e^{im\phi}
\approx k^{(m)}_n\,\J_m'(k^{(m)}_nd)e^{im\phi}
\]
\begin{equation}
\left.\frac{\partial_r\varphi}\varphi\right|_{(d,\phi)}\approx
\frac{k^{(m)}_n}{(\beta+i\gamma)d}
\label{bcA_}
\end{equation}
\paragraph{Region B (outside the opening of pinholes): $|r|\ge d$}
The wave equation is written as
\[
 -\{k_0^2-(1-i\epsilon)k_z^2\}\varphi(r,\phi)
=(1-i\epsilon)
\left(\partial_r^2+\frac 1r\partial_r
+\frac 1{r^2}\partial_{\phi}^2\right)\varphi(r,\phi)
\]
so that the solution is given by
\[
 \varphi\propto \H^{(1)}_m(k_Br)e^{im\phi}
\]
if we take into account its behavior when $r\rightarrow\infty$
($\H^{(1)}_m$ are the Hankel functions of the 1st kind).
Here $k_B$ satisfies $k_0^2=(1-i\epsilon)(k_B^2+k_z^2)$.
Noting that $k_B^2d^2-k_A^2d^2=\frac{i\epsilon}{1-i\epsilon}k_0^2d^2$
and that the absolute value of the right-hand side is much larger
that 1 (see (\ref{cp})) whereas $k_Ad$ in the left-hand side is of
order of 1 so that we can neglect this term and
(\ref{k_B2}) and (\ref{k_B}) holds as in the case of slit array.
From these we see that $|k_Bd|\gg 1$ and $\arg k_Bd\approx\frac\pi 4$
(so that $-\pi<\arg k_Bd<2\pi$), so that we can use the following
asymptotic form of $H^{(1)}_n$ for $\rho=k_Bd$ (see \cite{DLMF}):
\[
 \H^{(1)}_m(\rho)\approx\sqrt{\frac 2{\pi\rho}}
 \exp\left(i\left[\rho-\frac{2m+1}4\pi\right]\right)
\]
\[
 \H^{(1)\prime}_m(\rho)
 \approx i\sqrt{\frac 2{\pi\rho}}
 \exp\left(i\left[\rho-\frac{2m+1}4\pi\right]\right)
\]
so that finally we obtain
\begin{equation}
\left.\frac{\partial_r\varphi}\varphi\right|_{(d,\phi)}=ik_B.
\label{bcp}
\end{equation}
\paragraph{Boundary condition at $r=d$}
Continuity of $\varphi(r,\phi)$ and $\partial_r\varphi(r,\phi)$
at $r=d$ yields, from (\ref{bcA_}) and (\ref{bcp}),
\begin{equation}
ik_B
=\frac{k^{(m)}_n}{(\beta+i\gamma)d}.
\label{bcp_}
\end{equation}
(\ref{bcp_}) has the same form as (\ref{bcs}) with $k_n$ replaced
by $k^{(m)}_n$, so that similar to the case of slit array,
we obtain
\[
 k_A=k^{(m)}_n+\beta+i\gamma=\left(1-\frac{1+i}
{\sqrt{2\epsilon}k_0d}\right)k^{(m)}_n
\]
\[
 k_z=\sqrt{k_0^2-k_A^2}
\approx k_0-\frac{k^{(m)\,2}_n}{2k_0}+\frac{k^{(m)\,2}_n}
{\sqrt{2\epsilon}k_0^2d}i
\]
From this the attenuation of light along the waveguide
\[
 |e^{ik_zz}|^2=e^{-\zeta z}
\]
is calculated giving the attenuation coefficient
$\zeta=2\Im k_z$ as
\begin{equation}
 \zeta
=\frac{\sqrt 2k^{(m)\,2}_n}{\sqrt{\epsilon}k_0^2d}
=\frac{\sqrt 2\rho^{(m)\,2}_n}{\sqrt{\epsilon}k_0^2d^3}
=\frac{\sqrt{2L}\rho^{(m)\,2}_n}{\sqrt{\xi k_0^3}d^3}
\label{zeta}
\end{equation}
and the attenuation per length $L$ (i.e. per one
pinhole) $\zeta L$ is given by
\[
 \zeta L
=\frac{\sqrt{2}\rho^{(m)\,2}_n}{\sqrt{\xi}}
\left(\frac L{k_0d^2}\right)^{\frac 32}
=\frac{\rho^{(m)\,2}_n}{2\sqrt{\xi\pi^3}}
\left(\frac{\lambda L}{d^2}\right)^{\frac 32}.
\]

\subsection{Attenuation of a multi-transverse-mode light}
In the previous section we estimated the attenuation of a
single transverse mode wave. Each transverse mode is specified
by a pair $(m,n)$ of an integer $m$ and a non-negative integer $n$,
and if the wave is confined tightly enough in the pinhole waveguide,
the transverse mode functions are given by
\[
 \varphi_{mn}(r,\phi)=\left\{
\begin{array}{ll}
 \alpha_{mn}\J_m(k^{(m)}_nr)e^{im\phi}&(r\le d)\\
 0&(r>d)
\end{array}
\right.
\]
($\alpha_{mn}$ are the normalization factors).
Note that these are the Bessel beam transverse mode functions
clipped at one of their nodes
in the radial direction.
The orthonormal condition is written as
\[
 \delta_{mm'}\delta_{nn'}
 =\langle\varphi_{mn},\varphi_{m'n'}\rangle
 =\int_0^\infty rdr\int_0^{2\pi}d\phi\ \varphi^*_{mn}(r,\phi)
  \varphi_{m'n'}(r,\phi).
\]
In this section we consider, as an example, the case where a plane wave
$\psi_p=\frac 1{\sqrt{\pi d^2}}e^{ik_0z}$ is incident into the waveguide,
and calculate how the wave attenuates
while it propagates along the waveguide.
The incident wavefront is cut out at the input end of the waveguide
(we take the input end as $z=0$) giving the transverse wave function as
\[
 \varphi_p(r,\phi)=\frac 1{\sqrt{\pi d^2}}\theta(d^2-r^2)
\]
and such wavefront is, from the symmetry consideration, expanded
with only $m=0$ modes:
\begin{equation}
 \varphi_p=\sum_{n=0}^\infty\beta_n\varphi_{0n}
\label{beta}
\end{equation}
By integrating the square of absolute value of both sides of
the above equation in
$(r,\phi)$ plane, we see that $\sum_{n=0}^\infty
|\beta_n|^2=1$. The power attenuation is given by
\begin{equation}
P(z)=\sum_{n=0}^\infty|\beta_n|^2\exp(-\zeta^{(0)}_nz).
\label{Pz}
\end{equation}
Here $\zeta^{(m)}_n$ is $\zeta$ given in (\ref{zeta}).
By taking inner product of both sides of (\ref{beta}) with
$\varphi_{0n}$,
\[
 \beta_n=\langle\varphi_{0n},\varphi_p\rangle=
2\pi\alpha_{0n}^*\frac 1{\sqrt{\pi d^2}}\int_0^d rdr J_0(k^{(0)}_nr)=
\frac{2\sqrt\pi\alpha_{0n}^*}{k^{(0)}_n}\J_1(\rho^{(0)}_n)
=\frac 2{\rho^{(0)}_n}
\]
Here we used $\frac d{d\rho}(\rho\J_1(\rho))=\rho\J_0(\rho)$
and the value of $\alpha_{0n}$ derived
in the next subsection (\ref{al}).

\subsection{Normalization factors $\alpha_{0n}$}
From the normalization conditions of $\varphi_{0n}$ we find
\[
 1=\langle\varphi_{0n},\varphi_{0n}\rangle
 =2\pi|\alpha_{0n}|^2\int_0^d rdr\J_0^2(k^{(m)}_nr)
 =\pi|\alpha_{0n}|^2 d^2\J_1^2(\rho^{(0)}_n).
\]
Here we used the formula
\[
 \frac d{d\rho}\left\{\rho^2\frac{\J_0(\rho)^2+\J_1(\rho)^2}2\right\}
 =2\rho \J_0(\rho)^2
\]
and $\J_0(\rho^{(0)}_n)=0$.
$\alpha_{0n}$ are determined as, besides the phase factor,
\begin{equation}
 \alpha_{0n}=\frac 1{\sqrt\pi d\J_1(\rho^{(0)}_n)}.
\label{al}
\end{equation}

\end{document}